# Strategies for targeting chondrosarcomas *in vivo* and molecular dissection of oncogenic events in chondrosarcomas: is epigenetics the culprit?

Daoudi Rédoane*

*E-mail : redoane.daoudi@unicaen.fr – University of Caen Normandy 14 000 Caen France

* Personal email address : red.daoudi@laposte.net

**Abstract**

It is obvious that both epigenetic and non-epigenetic actors contribute to tumorigenesis in chondrosarcomas and more generally in other cancers. Thus, the main altered pathways in chondrosarcomas are now well established and include both epigenetic and non-epigenetic pathways such as the PI3K-AKT signaling, EGFR overexpression, SPARC overexpression, c-myc overexpression, IHH/GLI1 axis, loss of Rb function, HIF1-alpha stabilization, IDH1 mutations, hypermethylation and SIRT1. This review aims to provide a detailed analysis of these pathways and highlights recurrent interactions between non-epigenetic and epigenetic actors in chondrosarcomas, raising the intriguing possibility of developing therapeutics targeting both epigenetic and non-epigenetic actors and supporting data from previous studies. Finally, we propose some strategies for targeting chondrosarcomas in vivo based on properties of this tumor.

# Background

Chondrosarcomas are malignant cartilaginous sarcomas that are chemoresistant and radioresistant. Although chondrosarcomas are rare, they are potentially lethal tumors because they can spread to another part of the body like lungs. Epigenetic and non-epigenetic pathways are involved in the tumorigenesis of chondrosarcomas. Oncogenic events in chondrosarcomas mainly include cell proliferation, cell cycle progression, cell migration, cell survival, chemoresistance, radioresistance, angiogenesis and epithelial to mesenchymal transition. The latter appears paradoxical because sarcomas are, by definition, mesenchymal *ab initio*. However, it is thought that sarcomas can undergo EMT-related processes and display a biphenotypic morphology with properties of both mesenchymal (vimentin) and epithelial tumors (E cadherin expression)[1]. Hence, sarcomas can become either more mesenchymal (epithelial to mesenchymal transition) or epithelial (mesenchymal to epithelial transition). Here we describe the main epigenetic and non-epigenetic pathways involved in these oncogenic events in chondrosarcomas with regard to their interactions. As expected, we show that both epigenetic and non-epigenetic actors could constitute an interesting therapeutic target in chondrosarcomas and we summarize the involved pathways in this tumor. In the second part of this review, we describe some strategies for targeting chondrosarcomas *in vivo* based on properties of this tumor.

# Oncogenic pathways in chondrosarcoma: the role of epigenetics

## AKT-related pathways and oncogenic events in chondrosarcoma

The class I phosphoinositide-3-kinase (PI3K) produces PtdIns (3,4,5)P3 (PIP3) from PtdIns (4,5)P2 (PIP2) by phosphorylating PIP2. PIP3 is a phospholipid that resides on the plasma membrane of cells and it activates downstream targets like the kinase Akt for example. PIP3 can be dephosphorylated by the tumor suppressor PTEN, frequently deregulated in human cancers. The pleckstrin homology domain (PH) is a protein domain of approximately 120 amino acids found in several proteins such as Akt or PDK1. This domain is able to bind phosphatidylinositol lipids within membranes like PIP3 or PIP2. When PIP3 is produced by the class I PI3K, PH domains of Akt and PDK1 bind to PIP3. PDK1 can activate Akt by phosphorylating it on threonine 308[2]. Another kinase called PDK2 can activate Akt by phosphorylating it on serine 473[2]. Then Akt (Akt1) regulates several downstream targets in cells and the expression of these targets depends on both cell types and cell context. By regulating its downstream targets Akt has been shown to be involved in many different cellular processes, such as regulation of glucose metabolism in insulin-response tissues, cell proliferation, cell survival, protein synthesis (via mTORC1), angiogenesis and tumorigenesis. In chondrosarcoma AKT appears to be a key master regulator of a wide range of oncogenic events including EMT, migration, proliferation, survival and angiogenesis. In this part, we explore the involvement of the AKT pathway in tumorigenesis in chondrosarcoma by describing the main AKT-related pathways implicated in chondrosarcoma. Moreover, we assess whether epigenetics plays a significant role in tumorigenesis depending on these different pathways in chondrosarcoma.

### The double face of EGFR in chondrosarcoma: epigenetic and genetic alterations

EGFR is highly expressed in a wide range of solid tumors, including chondrosarcoma[3–5], thereby contributing to tumor aggressiveness. However, there are multiple molecular mechanisms underlying this overexpression in chondrosarcoma. EGFR acts as an upstream activator of the AKT pathway by activating the Ras pathway. In his turn, Ras is able to activate the PI3K[6]. The PI3K can also be activated by interacting with phosphorylated tyrosine residues in the EGFR. Because AKT seems to be a major regulator of oncogenic events in chondrosarcoma and that several EGFR inhibitors are already in clinical trials, EGFR could be a relevant therapeutic target in chondrosarcoma. Here, we explore the main pathways that may lead to EGFR overexpression in chondrosarcoma since it is important to elucidate these pathways in order to propose adequate treatments.

Firstly, EGFR amplification occurs in a subset of chondrosarcoma[5] leading to EGFR overexpression and AKT-related pathways activation. Secondly, DZNep, an S-adenosyl-l-homocysteine hydrolase (SAHH) inhibitor leading to a global methyltransferases inhibition, decreases EGFR expression in various cancer cell types[7,8]. Moreover, the activating histone modifications H3K27Ac and H3K4me3 are associated with EGFR expression in gliomas[9]. This result suggests that DZNep could decrease EGFR expression by preventing H3K4 trimethylation. In chondrosarcoma our preliminary results show that DZNep reduces EGFR expression. It may reduce EGFR expression either indirectly or directly. DZNep acts directly if it prevents H3K4 trimethylation of the EGFR promoter. DZNep may act indirectly if it decreases the expression of an activator of the EGFR promoter and/or enhances the expression of an inhibitor of the EGFR promoter. Additionally, a study shows that HDAC inhibitors like vorinostat decrease EGFR expression in colorectal cancer cells[10]. Surprisingly, even in the presence of HDAC inhibitors, the histones on EGFR promoter are hypoacetylated. This result can be explained by the fact that HDAC inhibitors prevent the association of CBP, a histone acetyltransferase, on EGFR promoter. EGFR expression is reduced because SP1, a transcription factor that enhances EGFR expression, is acetylated in the presence of HDAC inhibitors and the acetylation reduces the activity of SP1. These results indicate that HDAC2/3 (HDAC I), which are the histone acetyltransferases interacting with SP1, deacetylate SP1 to enhance its transcriptional activity and subsequent EGFR expression in colorectal cancer cells. Moreover, HDAC I possess not only an epigenetic activity through histones deacetylation but also a non-epigenetic activity through interacting with other proteins like SP1 in colorectal cancer cells. This non-epigenetic activity of HDAC I on EGFR expression remains to be explored in chondrosarcoma. Because DZNep decreases EGFR expression in chondrosarcoma, EGFR is at least in part under epigenetic regulation. The role of acetylation on EGFR expression in chondrosarcoma is still unclear. EGFR also seems to depend on genetic factors because EGFR amplification is found in chondrosarcoma[5].

Furthermore, gefitinib, an EGFR tyrosine kinase inhibitor, induces growth arrest and inhibition of metastasis in JJ012, SW1353 and OUMS27 chondrosarcoma cell lines[4]. Gefitinib also decreases the expression of MMP2 and MMP9, that are known to be beta-catenin target genes. This result shows that the AKT pathway could be involved in MMP2 and MMP9 expression in chondrosarcoma. In fact, GSK3b is a direct downstream target of AKT and GSK3b inhibits beta-catenin by phosphorylating it and promoting its degradation via the E3 ubiquitin ligase component beta-TrCP[11]. AKT indirectly activates the beta-catenin pathway by inhibiting GSK3b[11]. As described previously, another evidence that the AKT pathway could be activated in chondrosarcoma is the fact that EGFR, highly expressed in chondrosarcoma, is an upstream activator of the AKT pathway. These data suggest that EGFR may be involved in EMT in chondrosarcoma because beta-catenin has been shown to be involved in both transcriptional repression of E-cadherin and transcriptional induction of EMT-related genes[12]. DZNep induces apoptosis and reduces cell migration in chondrosarcoma[13] but it remains unclear whether the cell death is caused by the decreased EGFR expression. In fact, gefitinib doesn't induce cell death in the previous study but it is possible that authors didn't assess this phenomenon. The effect of DZNep on EMT remains to be elucidated in chondrosarcoma but we speculate that DZNep may prevent EMT in chondrosarcoma because it decreases EGFR expression and EGFR may be involved in EMT in chondrosarcoma.

These findings reveal that EGFR could be an interesting therapeutic target in chondrosarcoma because its pharmacological inhibition reduces growth, metastasis and MMP expression. Even if an epigenetic modulation (HDAC inhibitors or methyltransferases inhibitors) seems to be sufficient to decrease EGFR expression, genetic factors like EGFR amplification should be taken into consideration during designing therapeutic strategies targeting EGFR in chondrosarcoma. Furthermore, HDAC I may possess a non-epigenetic activity through interacting with other proteins like SP1 to promote EGFR expression in chondrosarcoma. This result suggests that both epigenetic and non-epigenetic inhibitors may represent an interesting approach to treat chondrosarcoma because a close relationship between epigenetic and non-epigenetic actors may exist in chondrosarcoma. For instance, SP1 inhibition may alter SP1 activity and may prevent the interaction between the non-epigenetic actor SP1 and the epigenetic actor HDAC I in chondrosarcoma, inhibiting the non-epigenetic activity of HDAC I. Conversely, HDAC I inhibition not only may alter its epigenetic activity through histones deacetylation but may also inhibit SP1 activity in chondrosarcoma.

## SPARC as an inducer of the AKT signaling in chondrosarcoma?

In addition to EGFR-induced AKT activation, SPARC may be involved in AKT activation in chondrosarcoma. SPARC is overexpressed in chondrosarcoma[5] and it has been shown that SPARC mediates cell survival of gliomas by activating the AKT pathway[14]. Furthermore, SPARC also activates the AKT pathway in melanoma[15]. These data suggest that SPARC overexpression may be involved in AKT activation in chondrosarcoma and that it may work in concert with EGFR overexpression to induce oncogenic events related to AKT activation. On first glance, AKT appears to be a non-epigenetic actor that induces non-epigenetic pathways like cyclin D1 activation, protein synthesis via mTORC1 activation, c-myc activation. However, emerging evidence indicates that AKT is able to induce epigenetic pathways in chondrosarcoma. For example, the AKT pathway could induce EMT in human sacral chondrosarcoma by recruiting Snail[16]. In his turn snail interacts with the histone demethylase LSD1 to demethylate H3K4me3 of E-cadherin gene, resulting in loss of E-cadherin expression and subsequent EMT[17–19]. In breast cancer cells, AKT induces slug expression via HSF-1[20] but this pathway is not elucidated in chondrosarcoma. These results suggest that SPARC may induce not only non-epigenetic pathways, but also epigenetic pathways in chondrosarcoma.

These data provide a better understanding of the role of SPARC in chondrosarcoma and shows, once again, that epigenetic and non-epigenetic pathways appear to be closely linked, impacting targeted therapies.

## C-myc in chondrosarcoma

AKT may be able to induce proliferation in chondrosarcoma by activating the proto-oncogene c-myc. In fact, AKT phosphorylates and inactivates GSK3b[11] and in his turn GSK3b phosphorylates c-myc on threonine 58, resulting in its ubiquitin-mediated proteasomal degradation[21]. Additionally, Ras-activated Erks stabilize c-myc by phosphorylating it on serine 62[22] and the Ras pathway can be induced by EGFR activation. Interestingly EGFR is highly expressed in chondrosarcoma (see above). Furthermore, it is known that both c-myc amplification and c-myc overexpression occur in chondrosarcoma[23], a phenomenon that likely contributes to cell proliferation in this tumor.

C-myc may also be involved in Warburg effect in chondrosarcoma. Warburg effect is the phenomenon in which cancer cells perform lactate fermentation even in the presence of oxygen. Multiple actors are thought to be implicated in this phenomenon, including c-myc, lactate deshydrogenase, HIF1-alpha and PDK1. C-myc induces LDH-A expression[24] and stabilizes HIF1-alpha[25]. In his turn, HIF1-alpha is able to activate PDK1 and PDK1 inhibits pyruvate deshydrogenase[26]. Additionally, HIF1-alpha induces LDH-A expression in cancer cells[27]. Snail is a target gene for HIF1-alpha[28], suggesting that c-myc may also induce EMT in chondrosarcoma by activating snail via HIF1-alpha.

These findings also indicate that c-myc may act via non-epigenetic pathways and epigenetic pathways in chondrosarcoma.

## The hedgehog signaling pathway in chondrosarcoma

The IHH pathway is active in some chondrosarcomas. An inactivating SUFU mutation and a gli1 amplification is identified in some chondrosarcomas[29,30] whereas PTCH1 or SMO gene mutations are infrequent in chondrosarcoma[31]. Gli1 inhibition suppress cell growth, cell cycle progression and induces apoptosis in human chondrosarcoma cells[32]. A cross-talk between AKT and IHH pathways may exist in chondrosarcoma. Indeed, AKT inhibits PKA and in his turn PKA can inactivate Gli1 by phosphorylating it on threonine 374[33]. Because c-myc is a target gene for gli1[34], the activation of IHH pathway could be associated with an epigenetic regulation depending on snail activation.

Consequently, the IHH pathway may induce EMT and other related events in chondrosarcoma. This hypothesis is consistent with the fact that HH pathway promotes EMT in lung squamous cell carcinomas[35]. In this study, gli1 is reversely associated with E-cadherin expression, suggesting that gli1 may induce c-myc expression and then c-myc may activate snail via HIF1-alpha. Finally, snail may recruit LSD1 to repress E-cadherin expression. However, as described previously, the IHH pathway is active in only some chondrosarcomas like central chondrosarcomas. In high-grade peripheral chondrosarcoma, the IHH signaling is decreased[36,37].

### The Rb pathway in chondrosarcoma: a role for Rb in suppressing EMT?

Rb alterations are found in chondrosarcoma[29,38,39]. Rb loss of heterozygosity (Rb-LOH) occurs in high-grade chondrosarcomas[38]. However, Rb gene mutations are rare in chondrosarcoma[39]. Thus, chondrosarcomas express low levels of Rb. Because Rb binds to the E2F1 transcription factor, involved in DNA replication-related genes expression, it is considered a tumor suppressor. The cyclin D1-cdk4 and cyclin D1-cdk6 complexes phosphorylate Rb during G1 phase. Then cyclin E-cdk2 complex also phosphorylates Rb to allow the S-phase cell cycle progression by disrupting association between Rb and E2F1. Because AKT pathway is active in chondrosarcoma and that AKT induces cyclin D1 activation, it also inhibits Rb and induces the S-phase cell cycle progression in chondrosarcoma. Moreover, Rb-LOH is responsible for the cell cycle progression in chondrosarcoma.

Interestingly, another tumor suppressor function of Rb is proposed in a study that shows that Rb inhibits EMT in MCF10A human mammary epithelial cells[40]. In this work, knockdown of Rb decreases E-cadherin expression but significantly induces slug and ZEB-1 expression. Moreover, Rb binds to the E-cadherin promoter. Because Rb prevents Slug and ZEB-1 expression, two transcription factors that play a major role in the induction of EMT by interacting with epigenetic factors such as LSD1, Rb indirectly regulates downstream epigenetic pathways in MCF10A human mammary epithelial cells. A non-epigenetic pathway regulated by Rb includes E2F1 sequestration for example.

Nevertheless, it remains unclear whether Rb inhibits epigenetic pathways and EMT in chondrosarcoma. Reexpression of Rb or cyclin D1 inhibitors could be proposed as a therapeutic strategy in chondrosarcoma.

### Deacetylation pathways in chondrosarcoma: more than just an epigenetic regulation?

SIRT1 (HDAC III) is a NAD-dependent protein deacetylase that plays an important role in deacetylating histone and nonhistone proteins. SIRT1 induces EMT in human chondrosarcoma cell lines SW1353 and HS.819.T[41]. In addition, SIRT1 induces Twist expression, a transcription factor involved in EMT. However, the mechanisms linking SIRT1 to Twist expression are not described in this study. Interestingly, another study shows that ET-1 promotes EMT in chondrosarcomas by inhibiting miR-300, a microRNA targeting Twist, via the AMPK pathway[42]. Other studies report that SIRT1 deacetylates LKB1 and induces translocation of LKB1 from the nucleus to the cytoplasm where it activates AMPK[43,44]. Taken together, these results indicate that SIRT1 deacetylates LKB1 in chondrosarcoma. In his turn, LKB1 activates AMPK and AMPK induces Twist expression by repressing miR-300 in chondrosarcoma. Then, Twist induces EMT by recruiting the methyltransferase SET8, which mediates H4K20 monomethylation[12], a histone mark that is associated with repression at E cadherin promoters. These results don't contradict the tumor suppressor role of AMPK. In fact, although AMPK is often classified as a tumor suppressor since it inhibits mTORC1 and protein synthesis, it can also promote tumor progression through non-epigenetic mechanisms, depending on both cell context and tumor type[45–47]. For example, another work shows that AMPK is responsible for metastasis in human chondrosarcoma cells[48]. Finally, a study indicates that SIRT1 is able to induce AKT activation[49] but it is unclear whether this phenomenon occurs in chondrosarcoma.

Therefore, SIRT1 has not only an epigenetic activity by deacetylating histones, but also a non-epigenetic activity by interacting with the LKB1-AMPK axis in chondrosarcoma. Additionally, SIRT1 may have a non-epigenetic activity by activating the AKT pathway in chondrosarcoma but further studies are required to address the question.

## IDH1 mutations in chondrosarcoma: beyond epigenetics?

Isocitrate dehydrogenase (IDH) exists in three isoforms and catalyzes the oxidative decarboxylation of isocitrate, producing alpha-ketoglutarate and carbon dioxide. Since metabolic alterations constitute a hallmark of several cancers, IDH mutations are involved in tumorigenesis. Interestingly, IDH mutations occur in a wide range of malignancies, including chondrosarcoma (IDH1 R132H, IDH1 R132G, IDH1 R132C). The purpose of this part is to describe the role of mutated IDH1 in chondrosarcoma. When IDH is mutated (i.e IDH1 and/or IDH2) it has a gain of function to produce the oncometabolite 2-HG, structurally similar to alpha-ketoglutarate and that acts as a potent inhibitor of alpha-ketoglutarate-dependent reactions[50], including histone demethylation (JmJC domain containing histone demethylases), DNA demethylation process (TET enzymes[51]) and HIF1-alpha degradation. The first remark is that IDH mutations can be responsible for non-epigenetic events in cancer cells through HIF1-alpha stabilization (angiogenesis for example).

In HCT116 and MCF-10A cells IDH-related mutations are responsible for EMT[52]. Importantly, 2-HG accumulation produces the same effect. In these cells, this result indicates that EMT may be induced in an epigenetic manner through hypermethylation of both DNA and histones. The global hypermethylation may induce the EMT phenotype in these cells by silencing tumor suppressors (H3K9me3, H3K27me3) and/or by activating oncogenes (H3K4me1, H3K4me2, H3K4me3, H3K36me3, H3K79me3)[53,54]. Moreover, HIF1-alpha could contribute to the EMT phenotype in these cells. In fact 2-HG accumulation prevents the degradation of HIF1-alpha and snail is a target gene for HIF1-alpha[28]. In his turn the transcription factor snail recruits the histone demethylase LSD1[17], that is not inhibited by 2-HG accumulation, to repress E-cadherin expression, an epigenetic modulation that is believed to contribute largely to the EMT process in cancer cells.

For these reasons, mutated IDH inhibition appears to be a good strategy in order to suppress both EMT and mutated IDH-related oncogenic process in mutated IDH tumors. AGI-5198, a selective mutant IDH1 inhibitor, causes demethylation of histone H3K9me3, growth inhibition of IDH1-mutant glioma cells *in vitro* as well as IDH1-mutant glioma xenografts in mice[55]. In the same way, in chondrosarcoma cell lines JJ012 and HT1080, AGI-5198 prevents colony formation (inhibition of cell proliferation) and migration[56]. Furthermore AGI-5198 induces apoptosis, cell cycle arrest and decreases 2-HG levels in a dose-dependent manner in JJ012 cells. Another study shows that both mutated IDH and wild-type IDH primary chondrosarcomas have no difference in the levels of H3K4me3, H3K9me3, H3K27me3, 5-hydroxymethylcytosine (5-hmC) or 5-methylcytosine (5-mC)[57]. At first glance, these results seem to contradict the fact that IDH mutants lead to a global hypermethylation in cells. However, they can be explained because chondrosarcomas are known to have a hypoxic microenvironment. Because histone demethylases and TET enzymes need molecular oxygen to perform demethylation, it is not surprising that wild-type IDH primary chondrosarcomas also exhibit hypermethylation of histones and high levels of 5-mC.

In another study, AGI-5198 has no effect on cell viability, colony formation, cell migration and global methylation (DNA and histones) in JJ012 cells[58]. This is in contrast to results of the previous study using the same cell line[56]. Although tumorigenic properties of JJ012 cells are not affected by AGI-5198, it is interesting to note that mutated IDH1 inhibition reduces 2-HG levels in a dose-dependent manner in JJ012 cells, similarly to the previous study. Taken together, the results suggest that the differences observed between the two studies may be due to the use of different techniques (for migration for example) or different experimental conditions. Note that these two papers were almost simultaneously published in two different journals. In this study[58] cell lines with IDH1 mutations have hypermethylation of CpG islands, a phenotype known as the CpG Island Methylator Phenotype (CIMP) and IDH1 mutated enchondromas also exhibit this phenotype[59,60]. Although some exceptions exist[61], the methylation of DNA represses transcription and could lead to cancer progression in chondrosarcomas with the CIMP. The wild-type cell lines lack this phenotype[58], suggesting that hypoxia doesn't occur in cell lines *in vitro* in contrast to primary chondrosarcomas. A study shows that T-cell acute lymphoblastic leukemia associated with good prognosis lacks the CIMP[62]. This result indicates that chondrosarcomas without IDH mutations may define a similar subtype of tumors associated with good prognosis. In the previous study, mutated IDH1 inhibition fails to reduce the global methylation in JJ012 cell lines[58]. This result is very interesting and strongly suggests that the global hypermethylation (CIMP and hypermethylation of histones) occurs through mechanisms other than the inhibition of demethylases in JJ012 cells. We postulate that both histone methyltransferases and DNA methyltransferases have a high methylation activity and/or their recruitment is enhanced at several loci in JJ012 cells with IDH1 mutations. Because the hypermethylation is not observed in cells without IDH1 mutations[58], we hypothesize that this high methylation activity and/or enhanced recruitment is selective to chondrosarcoma cell lines with IDH1 mutations. Indeed, in a previous study, mutated IDH1 inhibition in glioma cells is sufficient to reduce the methylation of histones[55], indicating that a high methylation activity doesn't occur in these cells. Taken together, these conclusions suggest that mutated IDH1 acts in concert with other epigenetic mechanisms to establish a global hypermethylation in mutated IDH1 chondrosarcoma cells. Furthermore, these results suggest that mutations in IDH1 are not essential for tumor progression in chondrosarcomas but only in enchondromas wherein IDH1 mutations seem to be involved in the initiation of enchondromas. This is consistent with the fact that IDH1 mutations are often early events in the development of other cancers[63]. In enchondromas, we propose that the other epigenetic mechanisms (i.e that don't depend on mutated IDH1 like methyltransferases) responsible for the high levels of methylation in chondrosarcomas would be negligible. According to the previous study[55], we hypothesize that both mutated IDH1 glioma cells and mutated IDH1 enchondroma exhibit a relatively similar methylation pattern (i.e cells wherein only mutated IDH1 induces hypermethylation without the help of other epigenetic mechanisms like methyltransferases) because mutated IDH1 inhibition in glioma dramatically reduces the hypermethylation. Additionally, we propose that the epigenetic mechanisms leading to the hypermethylation in mutated IDH1 chondrosarcoma cells consist in an increased methylation activity and/or in an increased recruitment of methyltransferases at several loci. This hypothesis is consistent with the fact that chondrosarcoma cell lines express high levels of histone methyltransferases like EZH2[13]. However, both wild-type IDH1 and mutated IDH1 chondrosarcoma cell lines express high levels of histone methyltransferases. For this reason, the global hypermethylation in mutated IDH1 chondrosarcoma cell lines is likely due to an increased recruitment of methyltransferases at several loci rather than an increased methylation activity.

In a previous study, mutated IDH1 inhibition reduces colony formation, migration and 2-HG levels whereas it induces apoptosis and cell cycle arrest in JJ012 cells[56]. This is in contrast with the second study in which mutated IDH1 inhibition has no effect on colony formation, migration and global methylation[58]. According to our previous conclusions and because this is the same cell line in which IDH1 is mutated, we propose that the methyltransferases are responsible for the hypermethylation in JJ012 cells in the two studies. Therefore, we anticipate that mutated IDH1 inhibition with AGI-5198 doesn't affect the hypermethylation in JJ012 cells in the first study[56]. The authors say that this investigation is ongoing in their laboratory. Now the question is "if AGI-5198 doesn't affect the hypermethylation in JJ012 cells in the first study, why it reduces colony formation, migration, induces apoptosis and cell cycle arrest?" and the second question is "Why AGI-5198 doesn't affect these oncogenic events in the second study using the same cell line?" A moderate dose of AGI-5198 (i.e 10 µM) is not sufficient to impair colony formation in both JJ012 and HT1080 chondrosarcoma cell lines while 2-HG levels are drastically decreased[56]. However, a high dose of AGI-5198 (i.e 20 µM) effectively prevents colony formation in the cells. This result indicates that the effect of AGI-5198 on colony formation doesn't depend on demethylases reactivation in these cells. Rather, it could depend on either a non-epigenetic activity of mutated IDH1 or an epigenetic activity of mutated IDH1 that differs from demethylases inhibition and that would be selectively inhibited by high doses of AGI-5198. A non-epigenetic activity for mutated IDH1 has been proposed in two studies[55,64] but if it exists, it remains unknown. This non-epigenetic activity differs from HIF1-alpha stabilization induced by mutated IDH1 since 2-HG levels are dramatically decreased by 10 µM of AGI-5198 and colony formation (proliferation) is unchanged. At first, this result seems to be in contrast with the fact that hypoxia induces proliferation via Erk activation in several cancer cells[65]. Moreover, our preliminary results suggest that hypoxia induces proliferation in chondrosarcoma cell lines by activating the Erk pathway. Note that Erk is also able to activate HIF1-alpha[66] to mediate a positive feedback loop that drives cell proliferation under hypoxic conditions and/or mutated IDH1 context. Nevertheless, the previous result is not in contrast with these studies if we consider that mutated IDH1 inhibition and subsequent HIF1-alpha-induced proliferation suppression is replaced by other proliferation-related pathways in chondrosarcoma cell lines, like the Akt pathway previously described in this review. A second possible explanation is the Warburg effect. For example, HIF1-alpha may be overexpressed in chondrosarcoma cell lines so that HIF1-alpha can't be totally degraded by prolyl-hydroxylases resulting in HIF1-alpha accumulation and lactate fermentation under normoxic condition (Warburg effect). Consequently, even if 2-HG levels are decreased, HIF1-alpha is not totally degraded and the Erk pathway remains activated in chondrosarcoma cell lines. This hypothesis seems consistent with a study showing that HIF1-alpha is overexpressed in chondrosarcoma tissues compared with normal tissues[67]. However, HIF1-alpha overexpression in this study may be due to hypoxia and not to a mutation. In fact, another study reveals the presence of HIF1 binding site in the HIF1-alpha promoter[68], suggesting a positive autoregulation by HIF1-alpha itself when it is stabilized under hypoxic condition. For this reason, the mechanism responsible for HIF1-alpha stabilization upon mutated IDH1 inhibition could be different from mutation-induced HIF1-alpha overexpression. We speculate that c-myc may induce Warburg effect by stabilizing HIF1-alpha in chondrosarcoma since a study shows that c-myc is responsible for HIF1-alpha stabilization under normoxic condition in MCF7 and T47D cells[25]. Moreover, as described previously, c-myc amplification is found in chondrosarcoma[23]. Note that a study indicates that SW1353 chondrosarcoma cells don't exhibit a metabolic profile consistent with the Warburg effect but authors didn't work with other chondrosarcoma cell lines[69].

In the first study, AGI-5198 (20 µM) also induces cell cycle arrest and apoptosis in JJ012 cells but not in HT1080 cells[56]. However, low doses of AGI-5198 has no effect on these events in the two cell lines. This result strongly suggests that a non-epigenetic activity of mutated IDH1 and/or an epigenetic activity not related to demethylases inhibition prevents cell cycle arrest and apoptosis in JJ012 cells but not in HT1080 cells, highlighting the differences between the chondrosarcoma cell lines and the importance to use different cell lines in order to better mimic the tumor context observed *in vivo*. Finally, in the first study, both low and high doses of AGI-5198 (from 1 µM to 20 µM) reduce cell migration in the two cell lines. This result indicates that decreased 2-HG levels are sufficient to reduce cell migration. Consequently, demethylases reactivation may cause an epigenetic modulation leading to a reduced migration in the two cell lines. Our previous hypothesis stating that methyltransferases occur in mutated IDH1 chondrosarcoma cell lines to establish the hypermethylation doesn't contradict the fact that demethylases reactivation is sufficient to reduce migration in these cells. Indeed, although we observe a global hypermethylation in mutated IDH1 chondrosarcoma cell lines, it is possible that certain loci (migration-related genes) are weakly methylated and that mutated IDH1 inhibition is sufficient to completely demethylate these genes and to inhibit the expression of these genes. This implies that local demethylation exists in chondrosarcoma cell lines and that these migration-related genes are initially activated by histone methylation (H3K4me3, H3K36me3 or H3K79me3 for example). Additionally, AGI-5198 (20 µM) reduces migration more effectively than AGI-5198 (10 µM)[56]. At these two doses, 2-HG levels are drastically decreased and similar. So, the observed difference could be due to a non-epigenetic activity of mutated IDH1 in these cells.

Now we describe the second study in which the previous oncogenic events are unchanged upon mutated IDH1 inhibition and using the same cell line as the first study: JJ012 cells[58]. The colony formation is not altered by the treatment (AGI-5198). However, authors use a maximum concentration of 10 µM AGI-5198. According to the previous study, this concentration appears to be insufficient to prevent colony formation in JJ012 cells. Cell migration is unchanged with AGI-5198 and the concentration should be sufficient to reduce cell migration. We propose that the observed differences in cell migration between the two studies are caused by the different techniques used to assess migration in JJ012 cells.

EMT is not described in these two studies and it is unclear whether mutated IDH1 inhibition is able to prevent EMT in chondrosarcoma cell lines. AGI-5198 could constitute an interesting approach to inhibit EMT in chondrosarcoma cell lines and further studies are needed. Interestingly, mutated IDH1 can be associated with favourable prognosis in glioblastoma[70–72]. This result shows that mutated IDH1 inhibition is not always a good therapeutic strategy and that the role of mutated IDH1 in cancer cells merit deeper investigation. It may depend on both cell lines and cell context (i.e *in vivo, in vitro or the species*). We propose two hypotheses to explain this favourable prognosis. The first is that methylation may lead to tumor suppressors inhibition (H3K9me3, H3K27me3, DNA methylation) and oncogenes activation (H3K4me3, H3K36me3, H3K79me3) in mutated IDH1 cells of poor prognosis patients whereas it may lead to tumor suppressors activation and oncogenes inhibition in mutated IDH1 cells of favourable prognosis patients. This "epigenetic switch" may depend on both cell lines and cell context. However, this first hypothesis seems inconsistent with the fact that DZNep impairs glioblastoma cancer stem cell self-renewal *in vitro* and *in vivo*[73]. In fact, this result suggests that methylation initially activates oncogenes in mutated IDH1 glioblastoma cells of favourable prognosis. The second hypothesis is that methylation is unchanged between mutated IDH1 cells of favourable prognosis patients and mutated IDH1 cells of poor prognosis patients (the same group of genes is methylated), this is the recruitment of demethylases that may be different. Demethylases may be recruited to tumor suppressors loci (that are initially inactivated by methylation) and oncogenes loci (that are initially activated by methylation) in mutated IDH1 cells of poor prognosis patients. In mutated IDH1 cells of favourable prognosis patients, demethylases may be recruited to other tumor suppressors loci (that are initially activated by methylation) and to other oncogenes loci (that are initially inactivated by methylation). Although certain tumor suppressors loci are activated by methylation and certain oncogenes loci are inactivated by methylation, this second hypothesis doesn't contradict the previous anti-cancer role of DZNep in glioblastoma because it is possible that the majority of methylation leads to the activation of oncogenes and the inactivation of tumor suppressors. However, long-term survival in glioblastoma patients is weakly correlated to IDH1 mutation[74]. This can be explained by the fact that, even if mutated IDH1 prevents both tumor suppressors and oncogenes demethylation leading to the favourable prognosis, it also prevents HIF1-alpha degradation resulting in enhanced tumorigenesis. Moreover, a non-epigenetic activity of mutated IDH1 responsible for proliferation, as previously described in this part, may still exist in cells of favourable prognosis patients. Finally, mechanisms other than mutated IDH1 (and previously mentioned) are responsible for tumorigenesis in glioblastoma and more generally in cancer. Therefore, AGI-5198 should be used only in mutated IDH1 cells of poor prognosis patients in order to selectively reactivate demethylases in these cells. These conclusions suggest that further studies are required to determine the role of mutated IDH1 in chondrosarcoma *in vivo*, that could differ from *in vitro* studies. Nonetheless, mutated IDH1 inhibition could be insufficient in chondrosarcoma *in vivo* since hypoxia strongly occurs in a wide range of solid tumors, including chondrosarcoma. In fact, even if mutated IDH1 inhibition is thought to reactivate demethylases and to allow degradation of HIF1-alpha *in vitro*, hypoxia inhibits demethylases and allows stabilization of HIF1-alpha. Additionally, high doses of AGI-5198 (20 µM) are required to suppress the eventual non-epigenetic activity of mutated IDH1 responsible for proliferation in chondrosarcoma *in vitro* and a high dose could be toxic to patients *in vivo*. Because methyltransferases-induced hypermethylation may occur in mutated IDH1 chondrosarcoma cells, DZNep may be exploited for therapeutic applications for chondrosarcoma. It inhibits methyltransferases and may improve the efficiency of AGI-5198 by preventing the remethylation by methyltransferases. Thus, even if DZNep alone is capable of triggering apoptosis in chondrosarcoma[13], a co-treatment with DZNep and low doses of AGI-5198 may constitute an alternative treatment for mutated IDH1 chondrosarcoma of poor prognosis patients *in vivo*.

To conclude, mutated IDH1 may act not only via its well-known epigenetic effects in chondrosarcoma, but also by interfering with either non-epigenetic pathways or epigenetic pathways not related to demethylases inhibition. Furthermore, mutated IDH1 inhibition appears to be insufficient to treat chondrosarcoma that harbor IDH1 mutations linked to poor prognosis *in vivo*. Instead, a co-treatment should be used. Unfortunately, both the chemoresistance and hypoxia limit the efficiency of treatments in chondrosarcoma.

## Conclusion

We note that the AKT pathway is intimately linked to tumor progression in chondrosarcoma. AKT is a central oncogenic signaling that orchestrates tumor progression through other pathways such as hedgehog, Rb or c-myc. As expected, non-epigenetic actors can indirectly modulate epigenetic pathways. For example, AKT that is a non-epigenetic actor, induces an epigenetic regulation of E cadherin expression by activating snail, a transcription factor that interacts with LSD1 to repress E cadherin expression in chondrosarcoma. Therefore, several epigenetic and non-epigenetic pathways are closely linked in chondrosarcoma. Surprisingly, several epigenetic actors (SIRT1, mutated IDH1) have also a non-epigenetic activity by interacting with non-epigenetic actors (AMPK, HIF1-alpha and other unidentified pathways) in chondrosarcoma, raising the possibility of developing therapeutics targeting these different non-epigenetic actors in order to prevent the non-epigenetic activity of epigenetic actors in chondrosarcoma. For instance, AMPK inhibition may prevent SIRT1-induced AMPK activation, Twist activation and oncogenic pathways related to AMPK activation. Another example is the inhibition of HIF1-alpha that may prevent mutated IDH1-induced HIF1-alpha activation, HIF1-alpha-induced snail activation and oncogenic pathways related to HIF1-alpha like angiogenesis. Conversely, SIRT1 inhibition inhibits its epigenetic activity and SIRT1-induced AMPK activation. Mutated IDH1 inhibition prevents its epigenetic activity (demethylases inhibition), its association with HIF1-alpha and its potential non-epigenetic activities. The interplay between epigenetic and non-epigenetic actors in chondrosarcoma suggests that epigenetics is not the culprit but a culprit and that non-epigenetic actors may play a more significant role than we thought and therefore constitute rational therapeutic targets in chondrosarcoma.

Table 1. An overview of possible epigenetic and non-epigenetic targets in chondrosarcomas.

| Target | Role | Drug | Mechanism |
| --- | --- | --- | --- |
| **Epigenetic actors** | | | |
| **Mutated IDH1** | Demethylases inhibition (except LSD1 family) | AG-120 (Ivosidenib) | IDH1 mutant inhibitor |
| | | AG-881 | IDH1 and 2 mutant inhibitor |
| **SIRT1** | Deacetylation of histones and non-histone proteins | Selisistat | SIRT1 inhibitor |
| **Methyltransferases** | Hypermethylation | DZNep | Competitive inhibitor of SAHH. Accumulation of SAH that inhibits a broad range of methyltransferases |
| **Non-epigenetic actors** | | | |
| **EGFR** | Participates in tumorigenesis by being an upstream activator of several pathways such as Ras or PI3K/AKT pathways | Gefitinib | EGFR inhibitor |
| | | Erlotinib | EGFR inhibitor |
| | | Cetuximab | EGFR antibody |
| **PI3K/AKT signaling** | Proliferation, migration, survival, angiogenesis, EMT, activation of several downstream pathways involved in tumorigenesis | Perifosine | AKT inhibitor |
| **C-myc** | Proliferation, Warburg effect | 10058-F4 | Inhibition of the c-myc/max heterodimerization |
| **IHH/GLI1 axis** | Proliferation, Warburg effect via c-myc activation | GANT-61 | Gli1 and 2 inhibitor |
| **Rb-LOH** | Proliferation | Palbociclib | CDK4 and 6 inhibitor |

| HIF1-alpha | Proliferation via Erk activation, EMT, Warburg effect | Chetomin | Blocks the interaction of HIF1-alpha and HIF2-alpha with transcriptional co-activator p300 |
|---|---|---|---|
| AMPK | EMT | Dorsomorphin | AMPK inhibitor |

**Figure 1. The epigenetic and non-epigenetic actors involved in chondrosarcomas are closely linked.** The dashed lines represent pathways that remain to be elucidated in chondrosarcomas.

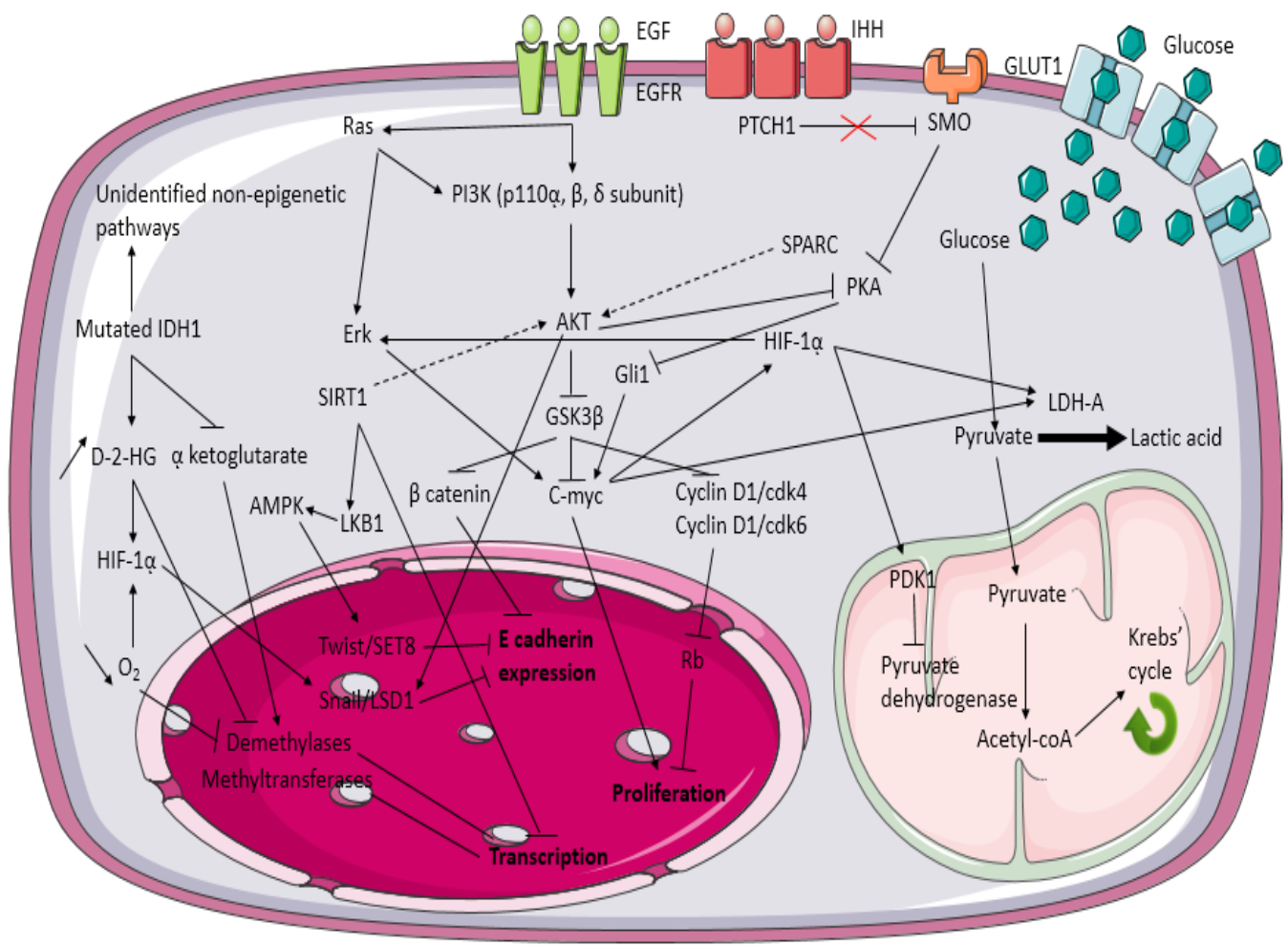

Now, we propose strategies that can be used in targeted drug delivery at the site of chondrosarcoma.

# Strategies for targeting chondrosarcoma: *in vivo* applications

Because chondrosarcomas exhibit specific tumor characteristics, it is possible to imagine drug delivery systems that can selectively target chondrosarcomas *in vivo*. Here we present strategies based on features of chondrosarcomas: hypoxia, acidosis, EGFR and survivin overexpression. However, note that most cancers share these characteristics with chondrosarcomas and that there are differences among cancer cells within a tumor, called intra-tumor heterogeneity. Consequently, chondrosarcoma cells produce lactic acid (hypoxia and/or Warburg effect) while others undergo aerobic metabolism through Krebs' cycle (normoxia).

### An orthotopic mouse model for human chondrosarcoma

Firstly, it is important to work with a functional mouse model. A study shows that mice injected with the chondrosarcoma cell line JJ012 develop tumors and lung micro-metastases whereas mice injected with FS090 cells don't develop tumors[75]. This mouse model replicates the site, morphology and many characteristics of human chondrosarcoma and can be used to test different strategies targeting chondrosarcoma *in vivo*.

### Acid sensitive linkers and hypoxia-activated prodrugs

Acidic extracellular pH is a property of tumors. The tumor microenvironment is generally more acidic than the normal tissues because hypoxia is frequently seen in solid tumors, including chondrosarcoma. As described previously, HIF1-alpha is induced and stabilized under hypoxic condition. It can be stabilized under normoxic condition (Warburg effect). Then, cancer cells produce lactic acid and an acidic microenvironment is formed. Acid sensitive linkers and hypoxia-activated prodrugs are useful to selectively deliver anticancer agents to chondrosarcoma and more generally to the cancer site[76]. The linker is attached to the prodrug and the prodrug is released at the cellular target site. The selection of a suitable linker is important since the linker may sometimes be inadequate, the linker and the prodrug may not combine, due to steric hindrance effect of the prodrug.

### Temperature-sensitive systems

*Background.* Another feature of tumors is that cancer cells have a higher temperature than the surrounding normal cells because their high metabolism generates heat. In fact, heat energy is released from cells during aerobic metabolism. At first, a thermosensitive polymer such as the poly(N-isopropylacrylamide) (poly-(NIPAAm)) appears to be a good strategy to improve drug delivery to chondrosarcoma. These polymers have a lower critical solution temperature (LCST), defined as the temperature at which the light transmission of the polymer solution drops to 90% of the original value. Below the LCST, the polymer is hydrophilic. Above the LCST, the polymer undergoes a phase transition to a hydrophobic state, generating turbidity and releasing the drug. For pure P-NIPAAm, the LCST is about 32°C but it can be tuned to anywhere in the range of 32°C to 50°C by incorporating acrylamide into the polymer chain. Moreover, any drug can be trapped in the polymer in contrast to the previous linker systems. Thus, thermosensitive polymers are interesting and function as on-off switches for drug release *in vivo*. However, other organs like the brain have a high metabolism and a higher temperature than other parts of the body. Additionally, moderate fever (39°C) is frequently seen in patients with cancer, including chondrosarcoma[77]. For these reasons, the LCST should be tuned to a value greater than 39-40°C for *in vivo* applications. Here we present one method for reaching this temperature within tumor: the technique is well documented and uses photothermal therapy.

*Photothermal therapy and chondrosarcoma.* It is possible to use gold nanocages covered with a thermosensitive polymer such as the P-NIPAAm (with a LCST tuned to a temperature greater than 40°C)[78,79]. Gold nanocages have a strong absorption in the near-infrared. When gold nanocages are exposed to near-infrared light, the light is absorbed and converted into heat through the photothermal effect. Gold nanocages containing the anticancer agent and covered with a thermosensitive polymer are injected intravenously. Then, gold nanocages are illuminated, generating local heating to directly kill cancer cells or indirectly by releasing the anticancer agent from the gold nanocages. Because EGFR is overexpressed in chondrosarcoma[3], it constitutes an interesting cell surface target and the previous system is improvable. Gold nanocages can be conjugated with anti-EGFR antibodies. The binding between EGFR and anti-EGFR antibodies initiates receptor-mediated endocytosis and leads to an increased concentration of gold nanocages inside the tumor.

However, although near-infrared window defines the range of wavelengths where light has its maximum depth of penetration in tissue, it exists a limited penetration depth of near-infrared light in tumors, reducing the effectiveness of the photothermal therapy. Because chondrosarcoma metastasizes most frequently to the lungs, it is important that near-infrared light reaches metastases or micro-metastases *in vivo*. For this reason, the treatment of metastases with the photothermal therapy appears to be ineffective and further studies are needed in order to know whether photothermal therapy can be used in chondrosarcoma, especially when there are not metastases yet.

# References

1. Sannino, G., Marchetto, A., Kirchner, T. & Grünewald, T. G. P. Epithelial-to-Mesenchymal and Mesenchymal-to-Epithelial Transition in Mesenchymal Tumors: A Paradox in Sarcomas? *Cancer Res.* **77,** 4556–4561 (2017).

2. Gold, M. R. *et al.* The B cell antigen receptor activates the Akt (protein kinase B)/glycogen synthase kinase-3 signaling pathway via phosphatidylinositol 3-kinase. *J. Immunol. Baltim. Md 1950* **163,** 1894–1905 (1999).

3. Song, Y. *et al.* Inhibition of EGFR-induced glucose metabolism sensitizes chondrosarcoma cells to cisplatin. *Tumour Biol. J. Int. Soc. Oncodevelopmental Biol. Med.* **35,** 7017–7024 (2014).

4. Song, J., Zhu, J., Zhao, Q. & Tian, B. Gefitinib causes growth arrest and inhibition of metastasis in human chondrosarcoma cells. *J. BUON Off. J. Balk. Union Oncol.* **20,** 894–901 (2015).

5. Movva, S. *et al.* Multi-platform profiling of over 2000 sarcomas: identification of biomarkers and novel therapeutic targets. *Oncotarget* **6,** 12234–12247 (2015).

6. Castellano, E. & Downward, J. RAS Interaction with PI3K: More Than Just Another Effector Pathway. *Genes Cancer* **2,** 261–274 (2011).

7. Lindsay, C. D. *et al.* Efficacy of EZH2 inhibitory drugs in human papillomavirus-positive and human papillomavirus-negative oropharyngeal squamous cell carcinomas. *Clin. Epigenetics* **9,** 95 (2017).

8. Takashina, T. *et al.* Combined inhibition of EZH2 and histone deacetylases as a potential epigenetic therapy for non-small-cell lung cancer cells. *Cancer Sci.* **107,** 955–962 (2016).

9. Erfani, P., Tome-Garcia, J., Canoll, P., Doetsch, F. & Tsankova, N. M. EGFR promoter exhibits dynamic histone modifications and binding of ASH2L and P300 in human germinal matrix and gliomas. *Epigenetics* **10,** 496–507 (2015).

10. Chou, C.-W., Wu, M.-S., Huang, W.-C. & Chen, C.-C. HDAC inhibition decreases the expression of EGFR in colorectal cancer cells. *PloS One* **6,** e18087 (2011).

11. Fang, D. *et al.* Phosphorylation of beta-catenin by AKT promotes beta-catenin transcriptional activity. *J. Biol. Chem.* **282,** 11221–11229 (2007).

12. Lamouille, S., Xu, J. & Derynck, R. Molecular mechanisms of epithelial-mesenchymal transition. *Nat. Rev. Mol. Cell Biol.* **15,** 178–196 (2014).

13. Girard, N. *et al.* 3-Deazaneplanocin A (DZNep), an inhibitor of the histone methyltransferase EZH2, induces apoptosis and reduces cell migration in chondrosarcoma cells. *PloS One* **9,** e98176 (2014).

14. Shi, Q. *et al.* Secreted protein acidic, rich in cysteine (SPARC), mediates cellular survival of gliomas through AKT activation. *J. Biol. Chem.* **279,** 52200–52209 (2004).

15. Fenouille, N. *et al.* SPARC functions as an anti-stress factor by inactivating p53 through Akt-mediated MDM2 phosphorylation to promote melanoma cell survival. *Oncogene* **30,** 4887–4900 (2011).

16. Yang, P. *et al.* SDF-1/CXCR4 signaling up-regulates survivin to regulate human sacral chondrosarcoma cell cycle and epithelial-mesenchymal transition via ERK and PI3K/AKT pathway. *Med. Oncol. Northwood Lond. Engl.* **32,** 377 (2015).

17. Ferrari-Amorotti, G. *et al.* Inhibiting interactions of lysine demethylase LSD1 with snail/slug blocks cancer cell invasion. *Cancer Res.* **73,** 235–245 (2013).

18. Li, Y. *et al.* LSD1-mediated epigenetic modification contributes to ovarian cancer cell migration and invasion. *Oncol. Rep.* **35,** 3586–3592 (2016).

19. Lin, Y. *et al.* The SNAG domain of Snail1 functions as a molecular hook for recruiting lysine-specific demethylase 1. *EMBO J.* **29,** 1803–1816 (2010).

20. Carpenter, R. L., Paw, I., Dewhirst, M. W. & Lo, H.-W. Akt phosphorylates and activates HSF-1 independent of heat shock, leading to Slug overexpression and epithelial-mesenchymal transition (EMT) of HER2-overexpressing breast cancer cells. *Oncogene* **34,** 546–557 (2015).

21. Gregory, M. A., Qi, Y. & Hann, S. R. Phosphorylation by glycogen synthase kinase-3 controls c-myc proteolysis and subnuclear localization. *J. Biol. Chem.* **278,** 51606–51612 (2003).

22. Escamilla-Powers, J. R. & Sears, R. C. A conserved pathway that controls c-Myc protein stability through opposing phosphorylation events occurs in yeast. *J. Biol. Chem.* **282,** 5432–5442 (2007).

23. Morrison, C. *et al.* MYC amplification and polysomy 8 in chondrosarcoma: array comparative genomic hybridization, fluorescent in situ hybridization, and association with outcome. *J. Clin. Oncol. Off. J. Am. Soc. Clin. Oncol.* **23,** 9369–9376 (2005).

24. Shim, H. *et al.* c-Myc transactivation of LDH-A: implications for tumor metabolism and growth. *Proc. Natl. Acad. Sci. U. S. A.* **94,** 6658–6663 (1997).

25. Doe, M. R., Ascano, J. M., Kaur, M. & Cole, M. D. Myc posttranscriptionally induces HIF1 protein and target gene expression in normal and cancer cells. *Cancer Res.* **72,** 949–957 (2012).

26. Kim, J., Tchernyshyov, I., Semenza, G. L. & Dang, C. V. HIF-1-mediated expression of pyruvate dehydrogenase kinase: a metabolic switch required for cellular adaptation to hypoxia. *Cell Metab.* **3,** 177–185 (2006).

27. Marín-Hernández, A., Gallardo-Pérez, J. C., Ralph, S. J., Rodríguez-Enríquez, S. & Moreno-Sánchez, R. HIF-1alpha modulates energy metabolism in cancer cells by inducing over-expression of specific glycolytic isoforms. *Mini Rev. Med. Chem.* **9,** 1084–1101 (2009).

28. Luo, D., Wang, J., Li, J. & Post, M. Mouse snail is a target gene for HIF. *Mol. Cancer Res. MCR* **9,** 234–245 (2011).

29. Tarpey, P. S. *et al.* Frequent mutation of the major cartilage collagen gene COL2A1 in chondrosarcoma. *Nat. Genet.* **45,** 923–926 (2013).

30. Tiet, T. D. *et al.* Constitutive hedgehog signaling in chondrosarcoma up-regulates tumor cell proliferation. *Am. J. Pathol.* **168,** 321–330 (2006).

31. Yan, T., Angelini, M., Alman, B. A., Andrulis, I. L. & Wunder, J. S. Patched-one or smoothened gene mutations are infrequent in chondrosarcoma. *Clin. Orthop.* **466,** 2184–2189 (2008).

32. Sun, Y. *et al.* Gli1 inhibition suppressed cell growth and cell cycle progression and induced apoptosis as well as autophagy depending on ERK1/2 activity in human chondrosarcoma cells. *Cell Death Dis.* **5,** e979 (2014).

33. Sheng, T., Chi, S., Zhang, X. & Xie, J. Regulation of Gli1 localization by the cAMP/protein kinase A signaling axis through a site near the nuclear localization signal. *J. Biol. Chem.* **281,** 9–12 (2006).

34. Ciucci, A. *et al.* Expression of the glioma-associated oncogene homolog 1 (gli1) in advanced serous ovarian cancer is associated with unfavorable overall survival. *PloS One* **8,** e60145 (2013).

35. Yue, D. *et al.* Hedgehog/Gli promotes epithelial-mesenchymal transition in lung squamous cell carcinomas. *J. Exp. Clin. Cancer Res. CR* **33,** 34 (2014).

36. Speetjens, F. M., de Jong, Y., Gelderblom, H. & Bovée, J. V. M. G. Molecular oncogenesis of chondrosarcoma: impact for targeted treatment. *Curr. Opin. Oncol.* **28,** 314–322 (2016).

37. Hameetman, L. *et al.* Peripheral chondrosarcoma progression is accompanied by decreased Indian Hedgehog signalling. *J. Pathol.* **209,** 501–511 (2006).

38. Röpke, M. *et al.* Rb-loss is associated with high malignancy in chondrosarcoma. *Oncol. Rep.* **15,** 89–95 (2006).

39. Yamaguchi, T. *et al.* Loss of heterozygosity and tumor suppressor gene mutations in chondrosarcomas. *Anticancer Res.* **16,** 2009–2015 (1996).

40. Arima, Y. *et al.* Rb depletion results in deregulation of E-cadherin and induction of cellular phenotypic changes that are characteristic of the epithelial-to-mesenchymal transition. *Cancer Res.* **68,** 5104–5112 (2008).

41. Feng, H. *et al.* The expression of SIRT1 regulates the metastaticplasticity of chondrosarcoma cells by inducing epithelial-mesenchymal transition. *Sci. Rep.* **7,** 41203 (2017).

42. Wu, M.-H. *et al.* Endothelin-1 promotes epithelial-mesenchymal transition in human chondrosarcoma cells by repressing miR-300. *Oncotarget* **7,** 70232–70246 (2016).

43. Price, N. L. *et al.* SIRT1 is required for AMPK activation and the beneficial effects of resveratrol on mitochondrial function. *Cell Metab.* **15,** 675–690 (2012).

44. Ruderman, N. B. *et al.* AMPK and SIRT1: a long-standing partnership? *Am. J. Physiol. Endocrinol. Metab.* **298,** E751-760 (2010).

45. Liang, J. & Mills, G. B. AMPK: a contextual oncogene or tumor suppressor? *Cancer Res.* **73,** 2929–2935 (2013).

46. Nakano, A. *et al.* AMPK controls the speed of microtubule polymerization and directional cell migration through CLIP-170 phosphorylation. *Nat. Cell Biol.* **12,** 583–590 (2010).

47. Vincent, E. E. *et al.* Differential effects of AMPK agonists on cell growth and metabolism. *Oncogene* **34,** 3627–3639 (2015).

48. Tsai, C.-H. *et al.* Resistin promotes tumor metastasis by down-regulation of miR-519d through the AMPK/p38 signaling pathway in human chondrosarcoma cells. *Oncotarget* **6,** 258–270 (2015).

49. Sundaresan, N. R. *et al.* The deacetylase SIRT1 promotes membrane localization and activation of Akt and PDK1 during tumorigenesis and cardiac hypertrophy. *Sci. Signal.* **4,** ra46 (2011).

50. Xu, W. *et al.* Oncometabolite 2-hydroxyglutarate is a competitive inhibitor of α-ketoglutarate-dependent dioxygenases. *Cancer Cell* **19,** 17–30 (2011).

51. Pastor, W. A., Aravind, L. & Rao, A. TETonic shift: biological roles of TET proteins in DNA demethylation and transcription. *Nat. Rev. Mol. Cell Biol.* **14,** 341–356 (2013).

52. Grassian, A. R. *et al.* Isocitrate dehydrogenase (IDH) mutations promote a reversible ZEB1/microRNA (miR)-200-dependent epithelial-mesenchymal transition (EMT). *J. Biol. Chem.* **287,** 42180–42194 (2012).

53. Pérez-Lluch, S., Guigó, R. & Corominas, M. Active transcription without histone modifications. *Oncotarget* **6,** 41401 (2015).

54. Pekowska, A., Benoukraf, T., Ferrier, P. & Spicuglia, S. A unique H3K4me2 profile marks tissue-specific gene regulation. *Genome Res.* **20,** 1493–1502 (2010).

55. Rohle, D. *et al.* An inhibitor of mutant IDH1 delays growth and promotes differentiation of glioma cells. *Science* **340,** 626–630 (2013).

56. Li, L. *et al.* Treatment with a Small Molecule Mutant IDH1 Inhibitor Suppresses Tumorigenic Activity and Decreases Production of the Oncometabolite 2-Hydroxyglutarate in Human Chondrosarcoma Cells. *PloS One* **10,** e0133813 (2015).

57. Cleven, A. H. G. *et al.* IDH1 or -2 mutations do not predict outcome and do not cause loss of 5-hydroxymethylcytosine or altered histone modifications in central chondrosarcomas. *Clin. Sarcoma Res.* **7,** (2017).

58. Suijker, J. *et al.* Inhibition of mutant IDH1 decreases D-2-HG levels without affecting tumorigenic properties of chondrosarcoma cell lines. *Oncotarget* **6,** 12505–12519 (2015).

59. Pansuriya, T. C. *et al.* Somatic mosaic IDH1 and IDH2 mutations are associated with enchondroma and spindle cell hemangioma in Ollier disease and Maffucci syndrome. *Nat. Genet.* **43,** 1256–1261 (2011).

60. Hughes, L. A. E. *et al.* The CpG island methylator phenotype: what's in a name? *Cancer Res.* **73,** 5858–5868 (2013).

61. Li, A. *et al.* Hypermethylation of ATP-binding cassette B1 (ABCB1) multidrug resistance 1 (MDR1) is associated with cisplatin resistance in the A549 lung adenocarcinoma cell line. *Int. J. Exp. Pathol.* **97,** 412–421 (2016).

62. Roman-Gomez, J. *et al.* Lack of CpG island methylator phenotype defines a clinical subtype of T-cell acute lymphoblastic leukemia associated with good prognosis. *J. Clin. Oncol. Off. J. Am. Soc. Clin. Oncol.* **23,** 7043–7049 (2005).

63. Watanabe, T., Nobusawa, S., Kleihues, P. & Ohgaki, H. IDH1 mutations are early events in the development of astrocytomas and oligodendrogliomas. *Am. J. Pathol.* **174,** 1149–1153 (2009).

64. Polychronidou, G. *et al.* Novel therapeutic approaches in chondrosarcoma. *Future Oncol. Lond. Engl.* **13,** 637–648 (2017).

65. Liu, L. *et al.* ERK/MAPK activation involves hypoxia-induced MGr1-Ag/37LRP expression and contributes to apoptosis resistance in gastric cancer. *Int. J. Cancer* **127,** 820–829 (2010).

66. Minet, E. *et al.* ERK activation upon hypoxia: involvement in HIF-1 activation. *FEBS Lett.* **468,** 53–58 (2000).

67. Chen, C. *et al.* Increased levels of hypoxia-inducible factor-1α are associated with Bcl-xL expression, tumor apoptosis, and clinical outcome in chondrosarcoma. *J. Orthop. Res. Off. Publ. Orthop. Res. Soc.* **29,** 143–151 (2011).

68. Pierre, C. C. *et al.* Methylation-dependent regulation of hypoxia inducible factor-1 alpha gene expression by the transcription factor Kaiso. *Biochim. Biophys. Acta* **1849,** 1432–1441 (2015).

69. Sanchez-Sanchez, A. M. *et al.* Melatonin Cytotoxicity Is Associated to Warburg Effect Inhibition in Ewing Sarcoma Cells. *PloS One* **10,** e0135420 (2015).

70. Zou, P. *et al.* IDH1/IDH2 Mutations Define the Prognosis and Molecular Profiles of Patients with Gliomas: A Meta-Analysis. *PLoS ONE* **8,** (2013).

71. Cheng, H.-B. *et al.* IDH1 mutation is associated with improved overall survival in patients with glioblastoma: a meta-analysis. *Tumour Biol. J. Int. Soc. Oncodevelopmental Biol. Med.* **34,** 3555–3559 (2013).

72. Hartmann, C. *et al.* Type and frequency of IDH1 and IDH2 mutations are related to astrocytic and oligodendroglial differentiation and age: a study of 1,010 diffuse gliomas. *Acta Neuropathol. (Berl.)* **118,** 469–474 (2009).

73. Suvà, M.-L. *et al.* EZH2 is essential for glioblastoma cancer stem cell maintenance. *Cancer Res.* **69,** 9211–9218 (2009).

74. Amelot, A. *et al.* IDH-Mutation Is a Weak Predictor of Long-Term Survival in Glioblastoma Patients. *PloS One* **10,** e0130596 (2015).

75. Clark, J. C., Akiyama, T., Dass, C. R. & Choong, P. F. New clinically relevant, orthotopic mouse models of human chondrosarcoma with spontaneous metastasis. *Cancer Cell Int.* **10,** 20 (2010).

76. Singh, Y., Palombo, M. & Sinko, P. J. Recent trends in targeted anticancer prodrug and conjugate design. *Curr. Med. Chem.* **15,** 1802–1826 (2008).

77. Nakamura, T., Matsumine, A., Matsubara, T., Asanuma, K. & Sudo, A. Neoplastic fever in patients with bone and soft tissue sarcoma. *Mol. Clin. Oncol.* **5,** 631–634 (2016).

78. Cobley, C. M., Au, L., Chen, J. & Xia, Y. Targeting gold nanocages to cancer cells for photothermal destruction and drug delivery. *Expert Opin. Drug Deliv.* **7,** 577–587 (2010).

79. Yavuz, M. S. *et al.* Gold nanocages covered by smart polymers for controlled release with near-infrared light. *Nat. Mater.* **8,** 935–939 (2009).